\documentstyle[aps,epsfig,multicol]{revtex}

\newcommand{\simle}
{\raisebox{-0.75ex}[-1.5ex]{$\;\stackrel{<}{\sim}\;$}}
\newcommand{\simge}
{\raisebox{-0.75ex}[-1.5ex]{$\;\stackrel{>}{\sim}\;$}}
\def\d{{\partial}}
\def\s{{\sigma}}
\def\e{{\epsilon}}
\def\k{{ {\bf k} }}
\def\p{{ {\bf p} }}
\def\q{{ {\bf q} }}
\def\Q{{ {\bf Q} }}

\def\w{{\omega}}
\def\a{{\alpha}}

\def\i{{ {\rm i} }}

\begin{document}
\draft

\def\runtitle{
Theory of anisotropic s-wave superconductivity 
with point-node like gap minima
}
\def\runauthor
 {Hiroshi {\sc Kontani}}

\title{
Theory of anisotropic s-wave superconductivity \\
with point-node like gap minima: \\
analysis of (Y,Lu)Ni$_2$B$_2$C 
}

\author{
Hiroshi {\sc Kontani}
}

\address{
Department of Physics, Saitama University,
255 Shimo-Okubo, Saitama-city, 338-8570, Japan.
}

\date{\today}

\maketitle      

\begin{abstract}
Recent intensive experimental studies revealed that
(Y,Lu)Ni$_2$B$_2$C is an anisotropic s-wave superconductor.
In addition, its gap function possesses deep point minima, 
whose ratio of the gap anisotropy is more than 10. 
On the theoretical side, however,
it is nontrivial to understand the origin of 
such a peculiar superconductivity.
In the present paper, we propose a mechanism
of the s-wave superconductivity with deep gap minima,
based on the theoretical model where 
strong electron-phonon coupling as well as
the moderate magnetic fluctuations coexist.
By analyzing the strong coupling Eliashberg equation,
we find that s-wave superconducting gap function
owing to the electron-phonon coupling becomes highly 
anisotropic as the magnetic fluctuations increases.
The set of model parameters
for realizing the strong gap anisotropy
in the present model
will be appropriate for (Y,Lu)Ni$_2$B$_2$C.
According to the present mechanism,
(groups of) pair of gap minima appear
at points on the Fermi surface which are
connected by the nesting vector ${\bf Q}$,
in both cases of s-wave superconductors
and non s-wave ones.
We briefly discuss 
other superconductors with highly anisotropic gap function, 
e.g.,
PrOs$_{4}$Sb$_{12}$ and $Na_{0.33}$CoO$_2$.

\end{abstract}



\begin{multicols}{2}
\section{Introduction}

\subsection{superconductivity in RENi$_2$B$_2$C}

Since the boron-carbide superconductors
RENi$_2$B$_2$C (RE=Lu, Y, Tm, Er, Ho, Dy)
was discovered about a decade ago
 \cite{Nature},
various experimental and theoretical studies 
have been devoted to determine the symmetry or
the mechanism of the superconductivity.
The crystal structure is body-centered tetragonal, 
consists of RE-C layers separated by Ni$_2$B$_2$ sheets.
Among them,
LuNi$_2$B$_2$C and YNi$_2$B$_2$C show relatively
high $T_{\rm c}$, 16.5K and 15.5K, respectively.
They are non-magnetic metals till very low temperatures.
On the other hand, 
in superconducting Er and Ho compounds,
$f$-electrons in Er and Ho show incommensurate
magnetic long range orders with $\Q_m\approx 2\pi(0.55/a,0,0)$,
where $a$ is the lattice spacing
 \cite{Er,Ho}.

Various experimental results revealed that 
LuNi$_2$B$_2$C and YNi$_2$B$_2$C 
are s-wave superconductors, whose
superconducting (SC) gap functions are highly anisotropic.
For example, 
the specific heat measurement in YNi$_2$B$_2$C 
below $T_{\rm c}$ tells that the SC gap in these compounds
changes from a gap-less type to a full-gap one,
by replacing Ni with Pt by 20\%
 \cite{Nohara}.
This result suggests that a s-wave superconductivity 
with deep point minima
occurs in a pure compound, and its anisotropy
is smeared out by impurities.
Later, Izawa et al. 
measured the $c$-axis thermal conductivity of YNi$_2$B$_2$C
in ${\bf H}$ rotated in various directions, 
and found that deep point minima in SC gap exist
along [100] and [010] directions
 \cite{Izawa}.
The estimated ratio of the anisotropy of the SC gap
will be more than 10.
Recently, 
Watanabe et al. confirmed the point-node like SC gap
function along [100] and [010] directions
by the ultrasonic attenuation measurement
 \cite{Watanabe}.

On the theoretical side, however,
the origin of such a highly anisotropic s-wave SC gap
has not been understood until now.
It cannot be reproduced by solving an Eliashberg equation
even if one assume a highly anisotropic dispersion
for conduction electrons 
and/or anisotropic electron-phonon (e-p) interactions.
Thus, a reasonable theoretical model for the
anisotropic s-wave SC gap with deep point minima,
whose ratio of anisotropy is more than 10, 
is highly demanded for understanding the superconductivity 
in (Y,Lu)Ni$_2$B$_2$C.

In the present paper,
we propose a mechanism of an
anisotropic s-wave superconductivity with deep gap minima,
in a system where
strong e-p coupling and moderate antiferromagnetic (AF) 
fluctuations coexist.
By solving the strong coupling Eliashberg equation,
we succeed in deriving a s-wave SC gap with deep point minima, 
whose ratio of anisotropy is more than 10, 
for a wider range of model parameters.
The point-node like SC gap in (Y,Lu)Ni$_2$B$_2$C observed 
experimentally is reasonably explained by the present theory.
The proposed mechanism of making deep gap minima
due to the magnetic fluctuations is simple and general, 
so it will also be realized 
in various superconductors, including the 
unconventional superconductors.
We study the case of the p-wave SC system in Appendix.

By various experimental and theoretical studies 
for (Y,Lu)Ni$_2$B$_2$C,
it is confirmed that both 
{\it (i) strong e-p couplings} and
{\it (ii) prominent AF fluctuations
due to the nesting of the Fermi surface (FS)}
coexist in these compounds.
Here, we explain the experimental and theoretical
evidences of (i) and (ii) in more detail.

\subsection{evidence for AF fluctuations}

NMR studies for (Y,Lu)Ni$_2$B$_2$C 
have been performed by several authors
 \cite{NMR,NMR-2}.
The spin-lattice relaxation ratio $1/T_1$ of $^{11}$B
and that of $^{89}$Y do not show Hebel-Slichter peaks.
In the normal state, on the other hand,
$1/T_1T$ of $^{11}$B increases monotonously
as temperature decreases, which suggests the
enhancement of the AF fluctuations at lower temperatures.
According to the spin fluctuation theory like the SCR theory
 \cite{Moriya},
$1/T_1T \propto \chi_Q^{2-d/2}$, where $d$ is the dimension
of the system and $\chi_Q$ is the staggered susceptibility.
In nearly AF metals,
$\chi_Q$ shows the Curie Weiss temperature dependence.
The observed $1/T_1T$ of $^{11}$B can be fitted well 
by the above expression both for d=2 and 3.

The strong AF fluctuations in 
(Y,Lu)Ni$_2$B$_2$C observed by NMR
are expected to originate from the nesting of the FS:
According to the LDA band calculations for LuNi$_2$B$_2$C
 \cite{band,band2,band3,band-nesting},
the FS possesses a nesting feature whose nesting vector is
$\Q \approx 2\pi(0.5/a,0), 2\pi(0,0.5/a)$.
The generalized susceptibility $\chi_\q^0(0)$
derived from the band structure given by the LDA study
shows a peak at $\q\sim \Q$
 \cite{band},
which means that the RPA-type magnetic susceptibility,
$\chi_\q(0)=\chi_\q^0(0)/(1-U\chi_\q^0(0))$,
has a sharp maximum at $\q\sim \Q$.

Another evidence of the nesting for $\q\sim \Q$
is the magnetic order of $f$-electrons
in Er and Ho compounds, whose ordering vector is 
$\Q_m\approx 2\pi(0.55/a,0,0)$.
This ordering is given by the RKKY interaction between 
$f$-electrons via the susceptibility of conduction electrons.
Because all the band structures for RENi$_2$B$_2$C
will be similar owing to the local nature of $f$-orbitals of RE,
$\chi_\q(0)$ in (Y,Lu)Ni$_2$B$_2$C is expected to
take the maximum value around $\q \approx \Q_m\approx \Q$.

\subsection{evidence for strong e-p coupling}

The density of states (DOS) 
at the Fermi energy in RENi$_2$B$_2$C 
obtained by the band calculation is 
$4.5\sim 4.8$(eV cell)$^{-1}$
 \cite{band2,band3,band-nesting}.
Because a cell contains two Ni atoms,
the DOS per Ni is $2.25\sim 2.4$eV$^{-1}$.
This value of the DOS is much larger than that 
for high-$T_{\rm c}$ cuprates, 
which is about $\sim 1.3$(eV cell)$^{-1}$
by the band calculation for La$_2$CuO$_2$,
where a cell contains one Cu atom.

Such a huge DOS in RENi$_2$B$_2$C 
suggests the strong e-p coupling.
In addition, the light weight of B atom
means the large frequency of the phonon mode, $\w_{\rm ph}$.
In fact, $\w_{\rm ph}\sim300$K is expected in these compounds
 \cite{band-nesting}.
The dimensionless e-p coupling constant $\lambda$ 
(or mass enhancement factor due to e-p interaction) 
is defined as $\lambda \equiv m^\ast/m_{\rm band}-1$,
where $m^\ast$ is the effective mass of an electron.
Its value can be estimated by the temperature dependence
of resistivity $\rho$ using the Bloch-Gruneisen
transport theory.
The experimental value $0.4\mu\Omega$cm/K
in LuNi$_2$B$_2$C leads to $\lambda\sim2.6$,
although it might be too overestimated
 \cite{band2}.
In addition, we stress that the prominent softening
of the phonon dispersion at $\k_{\rm obs}\sim \Q$ is observed
by neutron diffraction experiments
 \cite{Kohn}.
This result strongly suggests the strong e-p
coupling as well as the nesting of the FS with 
$\k_{\rm obs}\sim \Q$.
Thus, the observed Kohn anomaly
ensures the main character of the electronic properties
in (Y,Lu)Ni$_2$B$_2$C.

The estimations of the value of $\lambda$
using the thermodynamic measurements
have been tried by many authors
on the basis of the strong coupling Eliashberg equation.
References \cite{Michor} and \cite{Manalo}
concluded that $\lambda=1\sim 1.2$.
On the other hand,
$\lambda=0.5\sim0.8$ was deduced
using the scanning tunnelling spectroscopy
 \cite{tunneling}.
In addition,
a moderate momentum dependence of 
(or FS dependence of ) $\lambda$
was inferred be several authors:
Reference \cite{Hc2}
explained 
the experimental upper critical filed ($H_{\rm c2}$)
by assuming a two-band model with different $\lambda$'s
($\lambda_{\rm max}=0.8$ and $\lambda_{\rm min}=0.3$).
Later, more isotropic $\lambda$ was inferred by ref. 
 \cite{Manalo}.
More recently,
Yamauchi et al. studied the mass enhancement factor
$\lambda= m_{\rm dHvA}/m_{\rm band}-1$,
where $m_{\rm band}$ in the band mass given 
by the LDA study, and $m_{\rm dHvA}$
is the cyclotron mass measured by the dHvA study,
by assuming that the mass enhancement is caused 
only by e-p couplings
 \cite{Harima}.
The obtained $\lambda$'s are 0.1$\sim$0.76
depending on the portion of FS's.

However,
previous works based on the strong coupling study
could not reproduce the 
strongly anisotropic s-wave SC gap
as observed in (Y,Lu)Ni$_2$B$_2$C,
even if one assume a multi-band model with very 
different $\lambda$'s.
In the present paper,
we show the importance of the AF fluctuations
to reproduce the deep point minima of the gap function
found in (Y,Lu)Ni$_2$B$_2$C.

\subsection{s-wave or d-wave: theoretical viewpoint}

The existence of the prominent
AF fluctuations in RENi$_2$B$_2$C 
suggest the possibility of the d-wave SC state, 
like in high-$T_{\rm c}$ superconductors.
Fukazawa et al.
performed the third order perturbation analysis
with respect to $U$ in a two-dimensional tight-binding
Hubbard model for (Y,Lu)Ni$_2$B$_2$C, 
by neglecting the e-p couplings
 \cite{Fukazawa}.
By solving the Eliashberg equation, 
they found that $T_{\rm c}\sim 15$K
for d-wave symmetry is realized by choosing a 
reasonable strength of $U$.
However, the obtained $T_{\rm c}$ might be overestimated 
because d-wave $T_{\rm c}$ 
is relatively low in the case of 3D systems in general.
In fact, the FS of RENi$_2$B$_2$C
given by band calculations possesses
a three-dimensional (3D) structure,
rather than a two-dimensional one.
Experimentally, the anisotropy of the resistivity 
as well as that of $H_{\rm c2}$ are small.
In this sense, this compound is a 
three-dimensional superconductor.

Next, we crudely estimate the s-wave $T_{\rm c}$ 
due to the strong e-p coupling in RENi$_2$B$_2$C.
By taking strong coupling effects into account,
following BCS-McMillan 
 \cite{McMillan}
type expression for $T_{\rm c}$ valid for $\lambda\sim O(1)$ 
would be obtained:
\begin{eqnarray}
T_{\rm c}=(\w_{\rm ph}/1.2)\exp(-1/(\lambda^\ast-\mu^\ast)) ,
 \label{eqn:M}
\end{eqnarray}
where $\lambda^\ast \equiv \lambda/(1+\lambda)$
and $\mu^\ast$ is the Morel-Anderson pseudo-potential,
which represents the reduction of $T_{\rm c}$ 
owing to the pair-breaking effect by the Coulomb interaction.
By putting $\lambda=1.5$, $\w_{\rm ph}= 300$K, and
$\mu^\ast=0.15$ (a typical value) in eq.(\ref{eqn:M}),
we obtain $T_{\rm c}=27$K.
If we put $\lambda=1.0$ instead,
$T_{\rm c}=14$K is obtained.

In the following sections,
we study the s-wave superconductivity caused by the e-p coupling,
under the influence of prominent AF fluctuations.
We propose an effective model for (Y,Lu)Ni$_2$B$_2$C
which reflect their characteristic properties well.
By solving the strong coupling Eliashberg equation,
we succeed in deriving the s-wave SC gap function
with deep gap minima.
The proposed mechanism of the strongly anisotropic 
SC gap,
which is developed to explain the superconductivity 
in (Y,Lu)Ni$_2$B$_2$C,
might be applicable to several unconventional 
superconductors found recently.

\section{Theoretical Model}

\subsection{model with e-p coupling and AF fluctuations} 
In this section, we explain the theoretical model 
used in the present work.
In the present paper,
we study the mechanism of the 
anisotropic s-wave superconductivity
based on the simplified two-dimensional model
under the influence of the magnetic fluctuations,
which have not been studied previously.
The proposed mechanism is expected to occur
in real compounds which have complex three-dimensional FS's.

For the simplicity of the analysis, 
we assume a two dimensional isotropic Fermi surface
as shown in Fig.\ref{fig:FS}(a).
In addition, the density of states (DOS) at the Fermi level, 
$N(0)$, is assumed to be isotropic.
Hereafter, we put $k_{\rm F}=2.5$,
which corresponds to $n=0.50$ (quarter filling) 
if the area of the Brillouin zone is $(2\pi)^2$
(square lattice).

Next, we introduce the electron-electron interaction terms 
owing to the e-p interaction and the AF fluctuations.
The latter originates from
the (on-site) Coulomb interaction 
and the nesting of the FS.
First, we represent the interaction between electrons 
due to phonons, $V^{\rm ph}(\w)$.
In the case of the Einstein-type phonon, 
\begin{eqnarray}
V^{\rm ph}(\w+\i\delta)=
 \frac{g\w_{\rm ph}}{2}\frac{2\w_{\rm ph}}{(\w^2-\w_{\rm ph}^2)+i\w\delta},
 \label{eqn:EP-org}
\end{eqnarray}
where $\w_{\rm ph}$ is the frequency of the Einstein phonon.
$g$ has a dimension of energy, and
$\lambda\equiv gN(0)$ gives the dimensionless coupling constant
due to phonons.
In (Y,Lu)Ni$_2$B$_2$C,
$\w_{\rm ph}\sim300$K and $gN(0)\sim 1$ is expected
as mentioned in the previous section.

We also introduce the interaction between electrons
due to AF fluctuations, $V_\q^{\rm AF}(\w)$.
In a Hubbard model with on-site interaction, $U$,
it is given by $\frac{3U^2}{2}\chi_\q(\w)$
within the RPA or fluctuation-exchange (FLEX) type approximation
 \cite{Bickers}.
In the RPA or FLEX approximation,
the magnetic susceptibility $\chi_\q(\w)$ is given by 
$\chi_\q(\w)=\chi_\q^0(\w)/(1-U\chi_\q^0(\w))$,
where $\chi_\q^0(\w)$ is the irreducible susceptibility.
Here, we assume the following effective $V_\q^{\rm AF}$
whose validity is assured in a lower energy region:
\begin{eqnarray}
& &V_{\q}^{\rm AF}(\w+i\delta)
 \nonumber \\
& &\ \ \ \ \ 
 =\frac{a}{1+\xi_{\rm AF}^2|\Q-\q|^2 -i\w/\w_{\rm sf}}
 \ \ \ \ \ \ \ \ {\rm model \ 1} ,
 \label{eqn:model1} \\
& &\ \ \ \ \ 
 =\frac{2a}{(1+\xi_{\rm AF}^2|\Q-\q|^2)^2 -i\w/\w_{\rm sf}}
 \ \ \ \ \ {\rm model \ 2} ,
 \label{eqn:model2} 
\end{eqnarray}
where $\Q$ is the nesting vector where $\chi_{\q=\Q}(0)$
takes the maximum value, 
$\xi_{\rm AF}$ is the AF correlation length, and $\w_{\rm sf}$
represents the energy scale of the AF fluctuations. 
In both models, $a$
has a dimension of energy, so
$aN(0)$ represents the dimensionless coupling constant
for AF fluctuations.

The model 1 is derived directly from
$\chi_\q(\w)$ within the RPA or FLEX approximation, 
$\chi_\q(\w)=\chi_\q^0(\w)/(1-U\chi_\q^0(\w))$,
by expanding $U\chi_\q^0(\w)$ as
$U\chi_\Q^0(0)+ \a_{\rm S}\xi_{\rm AF}^2|\Q-\q|^2
-i\a_{\rm S}\w/\w_{\rm sf} +O(|\Q-\q|^4, \w^2)$,
where $\a_{\rm S}\equiv 1-U\chi_\Q(0) \ll 1$
is the Stoner factor.
Then, $a$ is given by $\frac{3U^2}{2}\chi_\Q(0)$.
Thus, model 1 is has a correct functional form 
when $|\q-\Q|\ll\xi_{\rm AF}^{-1}$ and $\w\ll\w_{\rm sf}$.
The model 1 had been studied intensively 
by Monthoux, Pines and their collaborators 
in the study of high-$T_{\rm c}$ superconductors
 \cite{Pines}.
Within the SCR theory 
 \cite{Moriya}
or FLEX-type approximation
 \cite{Bickers},
$\xi_{\rm AF}^2 \propto T^{-1}$ and
$a\propto\w_{\rm sf}^{-1}\propto \xi_{\rm AF}^2$,
which are observed experimentally
in various high-$T_{\rm c}$ cuprates.
Especially,
$\w_{\rm sf}>T$ ($<T$) in over-doped (under-doped)
cuprates above the pseudo-gap temperatures,
which is well reproduced by the FLEX approximation
 \cite{Kontani-Hall}.
Apparently, $\w_{\rm sf}$ becomes small in the close
vicinity of the magnetic quantum critical point (QCP).

However,
the model 1 might be unrealistic for
$|\q-\Q|\gg\xi_{\rm AF}^{-1}$ in that
it has a lorenzian form with respect to $|\q-\Q|$,
which decays too slow.
Instead of introducing a cutoff momentum $q_{\rm c}$
in the model 1,
we introduce the model 2 
which decays faster than lorenzian when 
$|\q-\Q|\gg\xi_{\rm AF}^{-1}$.
The functional form of the model 2 will be justified
if we assume the following expansion
$U\chi_\q^0(0)= U\chi_\Q^0(0)+ 2 \a_{\rm S} \xi_{\rm AF}^2|\Q-\q|^2
 + \a_{\rm S}^2 \xi_{\rm AF}^4|\Q-\q|^4 +O(|\Q-\q|^6)$,
where we fixed the coefficient of the forth order term
as $\a_{\rm S}^2 \xi_{\rm AF}^4$.
As a matter of convenience,
we put 2 in the numerator of eq.(\ref{eqn:model2})
to make the weights 
$\int_{-\infty}^\infty d^1q V_{\q}^{\rm AF}(0)$
for both models equivalent.
In the present study,
we treat $a$, $\xi_{\rm AF}$ and $\w_{\rm sf}$
as independent parameters
both in model 1 and in model 2.

The momentum dependences of model 1 and 2 
for $\w=0$ are shown in Fig. \ref{fig:bunpu},
where $x\equiv |\q-\Q|$. 
$x$-dependences of model 2 and model 3 (gaussian),
both of which decay faster than lorenzian,
looks similar.
Although model 1 have been studied intensively
in the study of high-$T_{\rm c}$ superconductors
 \cite{Pines},
its supremacy is not apparent for the purpose of the 
present study.
Note that a similar $T_{\rm c}$ for d-wave will
be obtained by solving the Eliashberg equation
in each models for the same parameters in 
$V_\q^{\rm AF}(\w)$.
In later sections, we study both model 1 and model 2,
and analyze the latter model mainly.

Finally, we point out that
the isotropic FS used in the present study
contradicts with the emergence of AF fluctuations,
which should arise as a result of the nesting of the FS.
This contradiction in the theoretical model
will be discussed later. 
Although this simplification is unrealistic,
it eliminates extrinsic complication in discussion
and make the result clear.
This simplification
will not harm the generality of the the mechanism of 
the superconductivity with deep gap minima 
proposed in the present paper.

\subsection{origin of the deep SC gap minima}

\begin{figure}
\begin{center}
\epsfig{file=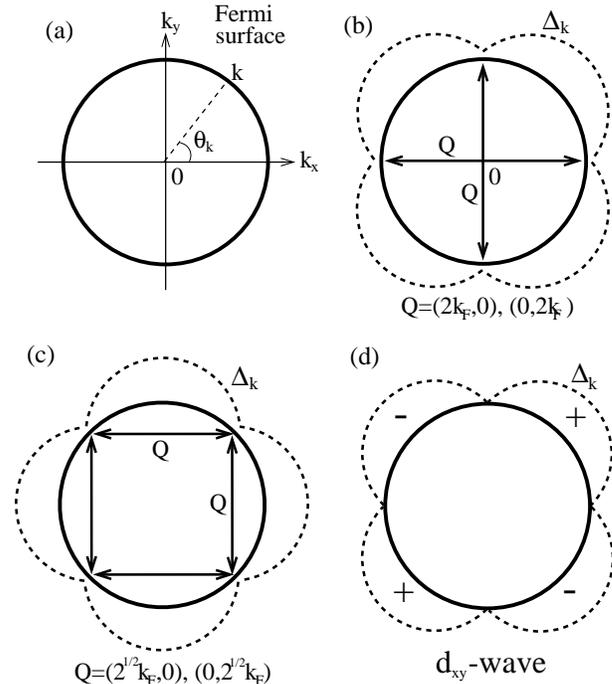,width=8cm}
\end{center}
\caption{
(a) Fermi surface of the present study. We put $k_{\rm F}=$,
which corresponds to the quarter-filled band in a square lattice
system.
(b) anisotropic s-wave SC gap 
in the presence of the AF fluctuations;
${\bf Q}=(2k_F,0),(0,2k_F)$.
(c) anisotropic s-wave SC gap 
in the case of ${\bf Q}=(\sqrt{2}k_F,0),(0,\sqrt{2}k_F)$.
(d) $d_{xy}$-wave SC gap
caused by the AF fluctuations;
${\bf Q}=(Q,0),(0,Q)$.
}
  \label{fig:FS}
\end{figure}

\begin{figure}
\begin{center}
\epsfig{file=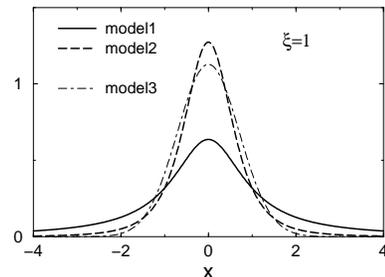,width=5cm}
\end{center}
\caption{
$V_{{\bf x}+\Q}^{\rm AF}(0)$ for model 1 and model 2,
with $a=1$ and $\xi_{\rm AF}=1$.
Model 3 is defined as
$V_{{\bf x}+\Q}^{\rm AF}(0)
 = \sqrt{\pi}a\cdot \exp(-\xi_{\rm AF}^2|{\bf x}|^2)$.
}
  \label{fig:bunpu}
\end{figure}

Here, we explain qualitatively about
the origin of the s-wave SC gap with deep minima.
In the weak coupling BCS theory,
where SC order parameter $\Delta_\k$ is treated as
energy-independent,
it will be allowed to replace eqs.(\ref{eqn:EP-org}) and 
(\ref{eqn:model1}) with $-g\theta(\w_{\rm ph}-|\w|)$
and $V_{\k-\p}^{\rm AF}(0)\cdot\theta(\w_{\rm sf}-|\w|)$,
respectively
 \cite{Schrieffer}.
As a result,
$T_{\rm c}$ within the weak coupling BCS theory
is given by the highest temperature
for the non-trivial solution $\Delta_\k\ne0$ of
\begin{eqnarray}
 \Delta_\k &=& \sum_\p g \frac{\Delta_\p}{2E_\p}
 {\rm tanh} \left(\frac{E_\p}{2T}\right)
 \theta(\w_{\rm ph}-E_\p)
 \nonumber \\
 & &- \sum_\p V_{\k-\p}^{\rm AF}(0)
 \frac{\Delta_\p}{2E_\p}
 {\rm tanh} \left(\frac{E_\p}{2T}\right)
 \theta(\w_{\rm sf}-E_\p) ,
 \label{eqn:weak-BCS}
\end{eqnarray}
where a singlet pairing is assumed.
In the above equation,
$E_\p \equiv \sqrt{(\e_\p-\mu)+\Delta_\p^2}$, where
$\e_\p$ is the dispersion of the conduction election and 
$\mu$ is the chemical potential.
The minus sign of the second term of eq.(\ref{eqn:weak-BCS})
can be interpreted as reflecting that
$\langle{\vec s}\cdot{\vec s}'\rangle=-3/4$ 
for a singlet Cooper pair.
In the absence of the AF fluctuations,
$T_{\rm c}^0=1.13\w_{\rm ph}\exp(-1/\lambda)$ 
for s-wave superconductivity is obtained by eq.(\ref{eqn:weak-BCS})
when $\lambda \equiv gN(0)\ll 1$
 \cite{Schrieffer}.
On the other hand, in the case of $g=0$,
the solution of eq.(\ref{eqn:weak-BCS})
is the $d_{xy}$-wave like owing to the second term of 
eq.(\ref{eqn:weak-BCS}),
$\Delta_{k_x,k_y}=-\Delta_{-k_x,k_y}
 =-\Delta_{k_x,-k_y}=\Delta_{-k_x,-k_y}$,
as shown in Fig. \ref{fig:FS}(d).
On the other hand,
$d_{x^2-y^2}$-type pairing is not favorable energetically
when ${\bf Q}\sim (2k_{\rm F},0), (0, 2k_{\rm F})$.

Here we consider the s-wave solution of the gap function
$\Delta_\k$ in eq.(\ref{eqn:weak-BCS})
owing to the strong e-p coupling.
As the coupling constant for
AF fluctuations, $aN(0)$, increases,
$\Delta_\k$ with s-wave symmetry
will has minima at $\theta_\k=n\pi/2$ ($n$ being a integer)
by reflecting the negative contribution from the 
second term of eq.(\ref{eqn:weak-BCS});
see Fig. \ref{fig:FS}(b).
Because of the simplicity of the mechanism,
this theory of the s-wave SC gap 
with deep minima due to AF fluctuations 
is expected to be general, independent of
the fine shape of the FS.
However, one has to check that
the s-wave $T_{\rm c}$, which will be smaller than 
$T_{\rm c}^0$ ($\equiv T_{\rm c}$ without AF fluctuations),
is larger than the d-wave $T_{\rm c}$ caused by AF fluctuations.

According to the present theory,
each point minima is connected with others by $\Q$.
In the case of $|\Q|<2k_{\rm F}$, 
eight point minima will emerge (instead of four),
except for $|\Q|=\sqrt{2}k_{\rm F}$ (see fig.\ref{fig:FS} (c)).
Except the number of point minima, 
the obtained results in the present study will qualitatively
hold even if $|\Q|<2k_{\rm F}$.
Here, we note that the position of the point gap minima 
is equivalent to that of ``hot spots'', where the quasiparticle 
damping rate due to AF fluctuations take the (local) maximum value.
The concept of the hot spot is important to understand
the anomalous transport phenomena in high-$T_{\rm c}$ cuprates
 \cite{Kontani-Hall}. 
When the AF fluctuations 
with ${\bf Q}=(Q,0), (0,Q)$ are strong enough,
a $d_{xy}$-type superconductivity 
as shown in fig.\ref{fig:FS} (d) will be realized
when $Q\simle 2k_{\rm F}$.
As shown in fig. \ref{fig:FS},
the position of the nodes for the s-wave state do not
coincide with that of the $d_{xy}$-wave state,
except for the case of $Q=2k_{\rm F}$.

In the present paper, we will confirm that
(i) a highly anisotropic s-wave SC gap 
observed in (Y,Lu)Ni$_2$B$_2$C 
(e.g., $\Delta_{\rm max}/\Delta_{\rm min}>10$)
can be realized with typical model parameters, and
(ii) a relatively higher s-wave $T_{\rm c}$
(e.g., $T_{\rm c} \simge T_{\rm c}^0/2$)
can be realized under the condition that s-wave $T_{\rm c}$ 
is larger than the d-wave one.
In the following sections,
we solve the Eliashberg equation
and derive the SC order parameter.
The obtained region of the model parameters 
where (i) and (ii) are satisfied
is wide enough and consistent with experimental situation 
in (Y,Lu)Ni$_2$B$_2$C.
Hereafter, we study only the case of 
$|{\bf Q}|=2k_{\rm F}$,
which corresponds to Fig.\ref{fig:FS} (b).

\section{Strong Coupling Eliashberg Equation}

In the preset section,
we solve the Eliashberg equation 
for a singlet paring, and
obtain the SC gap function and $T_{\rm c}$
for several sets of model parameters.
Because the coupling constant $\lambda\equiv gN(0)$ 
is of order $O(1)$
according to experiments,
which corresponds to the strong coupling superconductor,
we have to work on the strong coupling Eliashberg equation
by taking the energy dependence of the gap function,
$\Delta_\k(\w)$, into account.

The strong coupling Eliashberg equation for the present model
within the one-loop approximation
at zero temperature is given by
 \cite{Schrieffer,Allen}
\begin{eqnarray}
\Delta_\k(\w) &=&
 \frac1{Z_\k}N(0)\int_{\rm FS}\frac{d\Omega_\p}{2\pi}
 \int_{\Delta_\p^0}^\infty dz
 \nonumber \\
& & {\rm Re} \frac{\Delta_\p(z)}{\sqrt{z^2-\Delta_\p^2(z)}}\cdot
 \lambda_{\k-\p}(\w,z)
 \label{eqn:Eliashberg}
 \\
\lambda_{\k-\p}(\w,z) &=&
 \int_0^\infty dx \frac1\pi
 {\rm Im}\left(V^{\rm ph}(x)-V_{\k-\p}^{\rm AF}(x)\right)
 \nonumber \\
& &\times \left(\frac1{\w+z+x-i\delta}-\frac1{\w-z-x+i\delta}\right) 
 \label{eqn:Eliashberg2}
\end{eqnarray}
for s-wave SC state,
where $\Delta_\k^0$ gives the energy gap,
which satisfies that $\Delta_\k^0=\Delta_\k(\Delta_\k^0)$.
To simplify the discussion,
we neglect the pair breaking effect by
the Coulomb interaction $U$, that is,
$\mu^\ast\equiv N(0)U/(1+N(0)U{\rm ln}(W_{\rm band}/\w_{\rm ph}))=0$
is assumed
 \cite{Schrieffer,Allen}.
($\mu^\ast$ is often referred as the Morel-Anderson pseudo-potential.)
We also drop the impurity effect, that is, 
the clean limit case is studied.

In solving the Eliashberg equation at $T=0$,
we will not use $V^{\rm ph}(\w)$ given in eq.(\ref{eqn:EP-org})
because it gives an artificial singularity of $\Delta_\k(\w)$
at $\w=\w_{\rm ph}$ although it does not influence
$T_{\rm c}$ as well as $\Delta_\k^0$ badly.
To escape the singularity, we use 
\begin{eqnarray}
{\rm Im}V^{\rm ph}(\w+i\delta)=
\frac{g\w_{\rm ph}}{2}
 \frac{\Gamma}{(\w-\w_{\rm ph})^2+\Gamma^2} ,
 \label{eqn:EP}
\end{eqnarray}
where $\Gamma$ is a positive parameter 
which is much smaller than $\w_{\rm ph}$.
Equation (\ref{eqn:EP})
becomes $g\w_{\rm ph}(\pi/2)\delta(\w-\w_{\rm ph})$
if $\Gamma$ is infinitesimally small,
which is equivalent to eq.(\ref{eqn:EP-org}).
Hereafter, we put $\Gamma=\w_{\rm ph}/12$.

In eqs.(\ref{eqn:Eliashberg}) and (\ref{eqn:Eliashberg2}),
$V_{\k-\p}^{\rm AF}(x)$ is given 
in eq.(\ref{eqn:model1}) or eq.(\ref{eqn:model2}).
Hereafter, we put
$|{\bf Q}|=2k_{\rm F}$,
which corresponds to Fig.\ref{fig:FS} (b).
$Z_\k$ is the mass enhancement factor at $\w=0$,
which is given by
$Z_\k \equiv 1-\d\Sigma_\k(\w)/\d\w|_{\w=0}$,
where $\Sigma_\k(\w)$ is the normal self-energy
due to $V^{\rm ph}(\w)$ and $V_{\k-\p}^{\rm AF}(\w)$.
It is expressed at $T=0$ as
\begin{eqnarray}
Z_\k= 1+gN(0)+
 N(0)\int_{\rm FS}\frac{d\Omega_\p}{2\pi}
 V_{\k-\p}^{\rm AF}(0) .
 \label{eqn:Z}
\end{eqnarray}
Hereafter,
we numerically solve the set of Eliashberg equations,
eqs.(\ref{eqn:Eliashberg}), (\ref{eqn:Eliashberg2}) 
and (\ref{eqn:Z}),
under the condition that the gap function 
has the (anisotropic) s-wave type symmetry, i.e., 
$\Delta_\k(\w)$ has the four-fold rotational symmetry.

In deriving eqs.(\ref{eqn:Eliashberg})-(\ref{eqn:Z}),
we performed the integration with respect to $\e_\k$ first
by neglecting the imaginary part of the normal self-energy,
which represents the quasiparticle damping rate.
The validity of this approximation is apparently  
violated in high-$T_{\rm c}$ superconductors
because the large Im$\Sigma_\k(0) (\gg T)$ around the hot spots 
reduces $N(0)$.
This effect decreases $T_{\rm c}$ much.
The validity of this approximation
for the present model will be discussed later.


To obtain $T_{\rm c}$ both for s-wave and for d-wave,
we solve the following Eliashberg equation 
at finite temperatures 
 \cite{Allen}:
\begin{eqnarray}
\Delta_\k(\w_n) &=&
 -T\sum_{m}\frac{\pi N(0)}{Z_\k}
 \int_{\rm FS}\frac{d\Omega_\p}{2\pi}
 \frac{\Delta_\p(\w_m)}{\sqrt{\w_m^2+\Delta_\p^2(\w_m)}}
  \nonumber \\
& &\times \left(V^{\rm ph}(\w_n-\w_m)+V_{\k-\p}^{\rm AF}(\w_n-\w_m)
 \right) ,
 \label{eqn:Eliashberg-Tc} \\
V^{\rm ph}(\w_n)&=& -g\frac{\w_{\rm ph}^2}{\w_n^2+\w_{\rm ph}^2} ,
 \\
V_\q^{\rm AF}(\w_n)&=& 
 \frac{a\cdot 2^{\alpha-1}}
  {[1+\xi_{\rm AF}^2|\Q-\q|^2]^\alpha + |\w_n|/\w_{\rm sf}} 
 \label{eqn:VAF},
\end{eqnarray}
where $\w_n=\pi T(2n+1)$,
and $\alpha=1(2)$ for model 1(2).
$T_{\rm c}$ is given by the highest temperature
for the non-trivial solution of $\Delta_\k(\w_n)$,
under the constraint that
$\Delta_{k_x,k_y}=c\cdot\Delta_{-k_x,k_y}
 =c\cdot\Delta_{k_x,-k_y}=\Delta_{-k_x,-k_y}$,
where $c=1$ for the (extended) s-wave and
$c=-1$ for the d$_{xy}$-wave, respectively.
In deriving eq.(\ref{eqn:Eliashberg-Tc}), 
we used eq.(\ref{eqn:EP-org}), not eq.(\ref{eqn:EP}).

\section{Numerical Solutions}

\subsection{comparison between model 1 and model 2}

\begin{figure}
\begin{center}
\epsfig{file=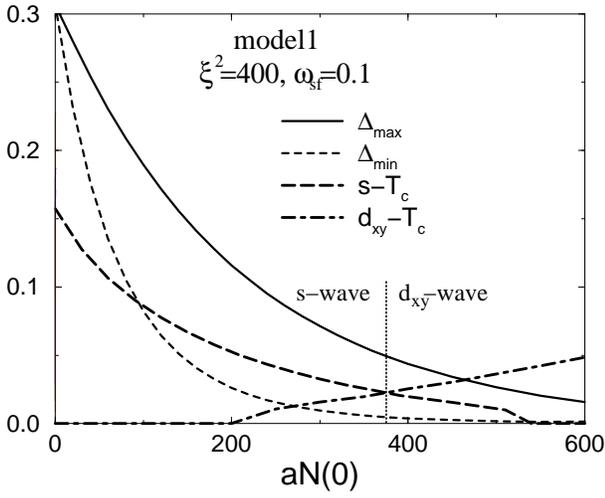,width=8cm}
\end{center}
\caption{$\Delta_{\rm min, max}$ and s,d-$T_{\rm c}$
obtained in model 1 as a function of $aN(0)$.
}
  \label{fig:Delta0-model1}
\end{figure}

Hereafter, we put the phonon parameters as
$\lambda\equiv gN(0)=1.5$ and $\w_{\rm ph}=1.0$;
the latter corresponds to $\sim300$K experimentally.
We stress that the main aim of this work
is to present the new mechanism of the point-node like 
SC gap, not to reproduce the precise experimental value of 
$T_{\rm c}$ and thermodynamic measurements in (Y,Lu)Ni$_2$B$_2$C.
Figure \ref{fig:Delta0-model1} shows
$\Delta_{\rm max}^0 \equiv \Delta_{\theta_k=\pi/4}^0$,
$\Delta_{\rm min}^0 \equiv \Delta_{\theta_k=0}^0$,
$T_{\rm c}$ for s-wave symmetry (s-$T_{\rm c}$) and
$T_{\rm c}$ for d-wave one (d-$T_{\rm c}$)
obtained in model 1
for $\xi_{\rm AF}^2=400$ and $aN(0)=0\sim600$.
We see that 
s-wave SC state is realized when $aN(0)<380$.
The condition $\Delta_{\rm max}^0/\Delta_{\rm min}^0>10$
is realized when $300<aN(0)<380$, 
although the realized $T_{\rm c}$ is less than
one fifth of $T_{\rm c}^0=0.157$,
which is the transition temperature without
AF fluctuations (i.e., $aN(0)=0$).

\begin{figure}
\begin{center}
\epsfig{file=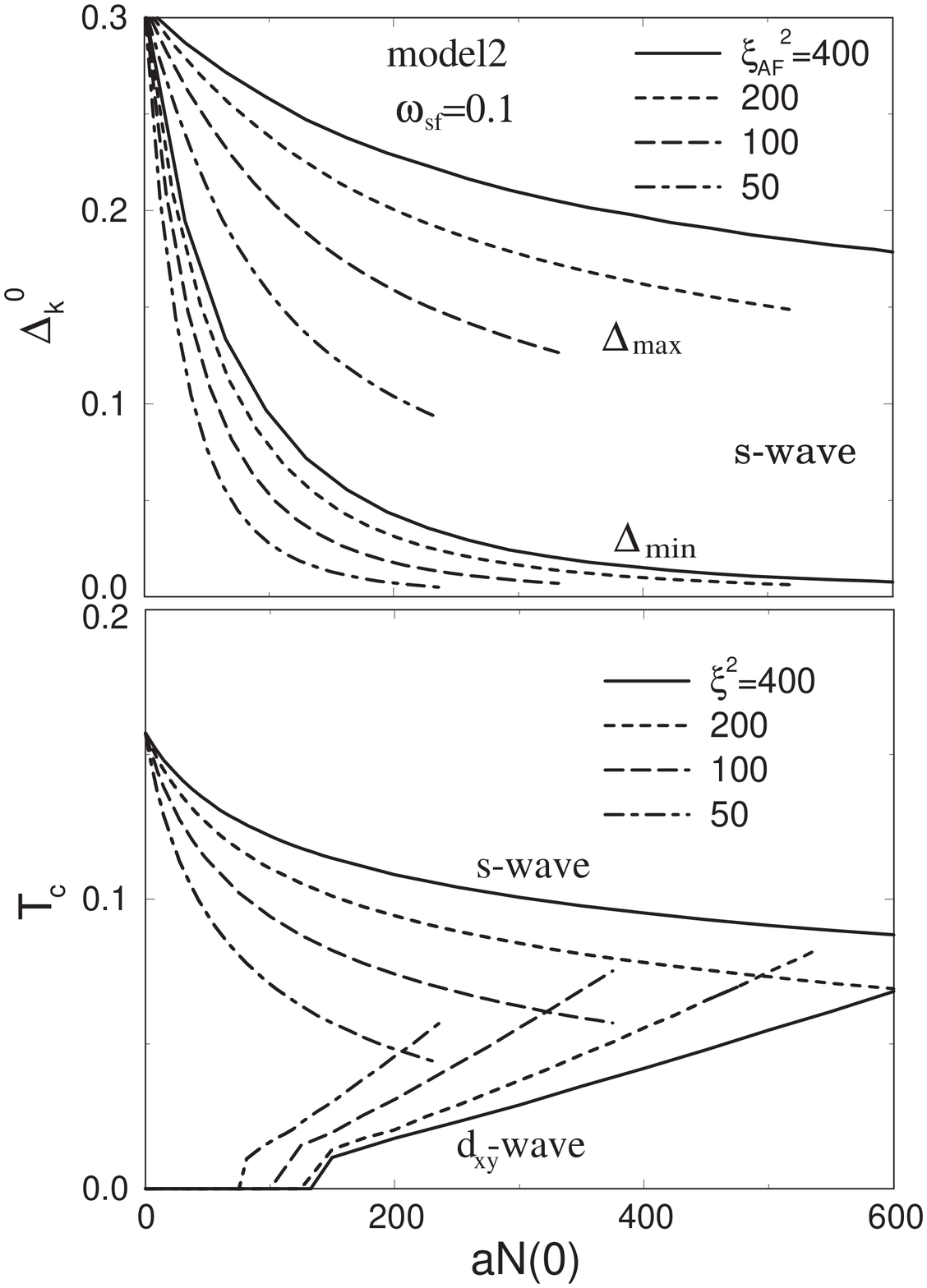,width=8cm}
\end{center}
\caption{
$\Delta_{\rm min, max}^0$ and s,d-$T_{\rm c}$
obtained in model 2 for $\w_{\rm sf}=0.1$.
}
  \label{fig:Delta0}
\end{figure}

Next, we study the model 2:
Figure \ref{fig:Delta0} shows 
$\Delta_{\rm max}^0$, $\Delta_{\rm min}^0$,
s-$T_{\rm c}$ and d-$T_{\rm c}$ for $\w_{\rm sf}=0.1$
in model 2.
We see that the condition 
$\Delta_{\rm max}^0/\Delta_{\rm min}^0>10$
is realized for much wider range of $aN(0)$
for $\xi_{\rm AF}^2=50\sim200$.
In addition, the realized s-wave $T_{\rm c}$
is higher than the half of $T_{\rm c}^0$.

As a result,
in both model 1 and model 2,
a highly anisotropic SC gap is realized 
under the condition of s-$T_{\rm c}>$d-$T_{\rm c}$.
However, 
the obtained s-wave $T_{\rm c}$ is 
much higher in model 2,
and the condition for a strong anisotropy
(say $\Delta_{\rm max}/\Delta_{\rm min}>10$)
is much easier.
The reason is that
the slow tail of model 1 for $\q$ away from $\Q$,
which follows the lorenzian form, 
does not contribute to make SC gap minima,
but work as a pair breaking effect 
like the momentum independent Coulomb interaction does.
(see Fig. \ref{fig:bunpu}.)

We note that
the lorenzian type slow tail of model 1
might not be natural because the peak of $\chi_\q(\w)$, 
which is created by the nesting of the FS, 
will decays faster once the nesting condition becomes ill
as $\q$ is away from $\Q$.
In reality, 
$\chi_\q(0)$ for high-$T_{\rm c}$ cuprates
obtained by the FLEX approximation,
which is shown in Fig. 11 of
 ref. \cite{Kontani-Hall} for example, 
seems to decay faster than the lorenzian form.
In that figure, we also see that 
$\chi_\q(\w)$ takes an almost constant value ($\sim 0.5U^{-1}$)
when $|\q-\Q|\gg\xi_{\rm AF}^{-1}$,
so it will work as a pair breaking effect.
As a result, the total
Morel-Anderson pseudo-potential will be
$\mu_{\rm AF}^\ast + \mu^\ast \sim 2\mu^\ast$,
where 
$\mu^\ast\equiv N(0)U/(1+N(0)U{\rm ln}(W_{\rm band}/\w_{\rm ph})$.
Hereafter, we study the condition of realizing 
the deep SC gap minima and analyze its property 
based on the model 2 in more detail,
by putting $\mu^\ast=0$.
The reader have to remind that
the obtained $T_{\rm c}$ in later sections
is over-estimated in that
the pair breaking effect is neglected.

\subsection{gap functions in model 2}

Figure \ref{fig:Delta0-2} shows
$\Delta_{\rm max}$, $\Delta_{\rm min}$,
s-$T_{\rm c}$ and d-$T_{\rm c}$
obtained in model 2 for $\xi_{\rm AF}^2=200$.
The $\w_{\rm sf}$-dependence of $\Delta_{\rm max,min}$
and s-$T_{\rm c}$ is very weak, while that of 
d-$T_{\rm c}$ is strong.
For $\w_{\rm sf}=0.1$,
$\Delta_{\rm max}/\Delta_{\rm min}>10$ is realized
for $aN(0)>280$, and 
the condition s-$T_{\rm c}>$d-$T_{\rm c}=0.075$ is 
satisfied when $aN(0)<490$.

Here, we discuss about the adequacy of parameters
used in the present calculation.
The relatively large value of $\xi_{\rm AF}$ used 
in the present analysis is consistent with 
the prominent AF fluctuations
in (Y,Lu)Ni$_2$B$_2$C at lower temperatures
 \cite{NMR,NMR-2}.
Then, the smaller value of $\w_{\rm sf}$ is also
expected because $\w_{\rm sf}\propto \xi_{\rm AF}^{-2}$.
Note that $\w_{\rm sf}\sim T$ and $\xi_{\rm AF}^2\propto T^2$
are realized in an optimum high-$T_{\rm c}$ cuprates, 
which is reproduced by the FLEX approximation
 \cite{Kontani-Hall}.
Moreover, the coupling constant for AF fluctuations,
$aN(0)$, which corresponds to $N(0)(3U^2/2)\chi_\Q(0)$
within the RPA or FLEX approximation,
will be smaller than that of high-$T_{\rm c}$ cuprates
because $U/W_{\rm band}$ in (Y,Lu)Ni$_2$B$_2$C is smaller.
As one can see in Fig. \ref{fig:Delta0-2},
similar strongly anisotropic gaps 
are obtained for wider range of parameters 
when $\xi_{\rm AF}^2=50\sim400$.

Figure \ref{fig:D0theta} 
shows the $\k$-dependence of $\Delta_\k^0$
on the Fermi surface, for $aN(0)=0\sim400$.
(Note that $\theta_\k={\rm tan}^{-1}(k_x/k_y)$.)
Thus, the s-wave SC gap with 
deep point minima
is realized for $aN(0)>300$.
It is similar to (s+g)-wave SC gap function
proposed in ref. 
 \cite{Maki},
where the impurity effect on the shape of $\Delta_\k^0$ 
is analyzed. 
We stress that a model with almost the same strength 
of attractive interactions
both for the s-wave and for the g-wave channels,
which was just introduced as an assumption in ref.
 \cite{Maki},
is derived microscopically in the present study.

Figure \ref{fig:Gap-w} shows the
$\e$-dependence of $\Delta_\k(\e)$.
Note that $\Delta_\k(\Delta_\k^0)=\Delta_\k^0$
gives the energy-gap of the SC order parameter.
We comment that 
$\Gamma=\w_{\rm ph}/6$ in eq.(\ref{eqn:EP})
is assumed only in deriving Fig. \ref{fig:Gap-w},
although $\Gamma=\w_{\rm ph}/12$ is used in other figures.
(By this reason, $\Delta_\k^0$ for $aN(0)=0$ is
smaller than 0.3 only in Fig. \ref{fig:Gap-w}.)
In Fig. \ref{fig:Gap-w},
we note that $\Delta_{\rm min}(\e)$ (for $\theta_\k=0$)
increases as $\e$ increases from $\Delta_{\rm min}^0$.
Such an energy dependence of $\Delta_{\rm min}(\e)$
will stabilize the solution for the strongly anisotropic
s-wave SC state in eq.(\ref{eqn:Eliashberg}).
This effect, which is dropped within the weak coupling BCS theory,
means the importance of the strong coupling effect
for realizing the strongly anisotropic SC gap.

At finite temperatures,
the gap function $\Delta_\k(\e_n)$ is obtained by 
solving the strong coupling Eliashberg equation 
for Matsubara frequencies, eq (\ref{eqn:Eliashberg-Tc}).
Then, the gap function for real $\e$,
$\Delta_\k(\e)$, is given by the numerical 
analytic continuation with high accuracy.
Here, we study several physical quantities
at finite temperatures
using the obtained $\Delta_\k(\e)$:
Figure \ref{fig:DOS} 
show the derived DOS for $\xi_{\rm AF}^2=200$ 
and $aN(0)=0\sim400$,
which is given by 
\begin{eqnarray}
\rho(\e)&=& \sum_\k{\rm Im}G_\k(\w-\delta)/\pi
 \nonumber \\
 &=& N(0)\int_{\rm FS}\frac{d\Omega_\k}{2\pi}
 {\rm Re}\frac{|\e|}{\sqrt{\e^2-\Delta_\k^2(\e)}} .
\end{eqnarray}
We put $N(0)=1$ in Fig. \ref{fig:DOS}.
It is shown that the finite DOS at lower energies
emerges as the coupling constant for AF fluctuations,
$aN(0)$, increases.
We can also calculate
the nuclear-lattice relaxation ratio ($T_1$)
and the ultrasonic attenuation ratio ($\a$).
Figure \ref{fig:T1T} 
shows the obtained $1/T_1T$ and $\a$,
which are given by
\begin{eqnarray}
R_{\pm} &=& N(0)\int_{\rm FS}\frac{d\Omega_\k}{2\pi}
 \int_0^\infty dz \left(-\frac{df}{dz}\right)
 \nonumber \\
& &\times 2\left([g_\k(z)]^2 \pm [f_\k(z)]^2 \right) ,
 \\
g_\k(z)&\equiv& \int d\e_\k {\rm Im}G_\k(z-i\delta)/\pi
 = {\rm Re}\frac{z}{\sqrt{z^2-\Delta_\k^2(z)}} ,
 \\
f_\k(z)&\equiv& \int d\e_\k {\rm Im}F_\k(z-i\delta)/\pi
 = {\rm Re}\frac{\Delta_\k(z)}{\sqrt{z^2-\Delta_\k^2(z)}} ,
 \label{eqn:Rpm}
\end{eqnarray}
where $1/T_1T=R_+$ and $\a=R_-$, respectively.
We note that $\a$ in eq. (\ref{eqn:Rpm})
is derived by taking the average of the
momentum of the absorbed phonon,
which might be different from the experimental situation.
In Fig. \ref{fig:T1T},
$1/T_1T$ for $aN(0)=0$
shows a large coherence peak below $T_{\rm c}$.
However, it is smaller than that given by the weak coupling
BCS theory in the present model
because the peak of the DOS is smeared out
due to the strong coupling effect,
that is, 
finite ${\rm Im}\Delta_\k(\e)$ at finite temperatures.
The coherence peak is further suppressed
as $aN(0)$ increases.

We comment that the coherence peak 
is further suppressed or vanishes 
if we take account of 
the effect of the finite quasiparticle damping rate,
$\gamma_\k={\rm Im}\Sigma_\k(-i\delta)$, at finite temperatures,
because finite $\gamma_\k$ reduce the DOS at $\e=0$.
However, this effect cannot be included in eq.(\ref{eqn:Eliashberg-Tc})
because the following relation:
\begin{eqnarray}
\int_{-\infty}^\infty d\e_\k {\hat G}_\k(\w_n)= -\pi
 \frac{i\w_n Z_\k{\hat 1} + \Delta_\k(\w_n){\hat \sigma}_x}
 {\sqrt{[\w_n Z_\k]^2+\Delta_\k^2(\w_n)}} ,
 \label{eqn:Gint}
\end{eqnarray}
which is given by dropping the 
self-energy except for its $\w_n$-linear term, $(-Z_\k+1)i\w_n$, 
has been used in deriving 
eq.(\ref{eqn:Eliashberg-Tc}).
In eq.(\ref{eqn:Gint}),
${\hat G}_\k(\w_n)$ is the $2\times2$ Green function
in Nambu representation, and ${\hat \sigma}_x$
is the Pauli matrix.
To take the effect of $\gamma_\k$ into account,
which has been very important in the case of 
high-$T_{\rm c}$ superconductors, 
we have to perform the $\e_\k$-integration ``numerically'' 
instead of using eq.(\ref{eqn:Gint}).
We further comment that 
the anisotropy of the dispersion for conduction electrons,
which is totally neglected in the present model,
will decreases the coherence peak 
because a small $\k$-dependence of $\Delta_\k$ emerges
even in the absence of the AF fluctuations.
By these two reasons mentioned above, 
the tiny coherence peak for $aN(0)>300$
may vanish after all.
Note that a tiny coherence peak exists even in the case of $d$-wave
within the BCS theory
if the momentum dependence of $N(0)$ is totally neglected.

Figure \ref{fig:T1T}
also show the log($1/T_1T$)-log$T$ plot 
for $aN(0)=0\sim400$.
This plot shows that
the temperature dependence of $1/T_1$ at lower temperatures
changes from an exponential behavior to a $T^{3}$-behavior
for $aN(0)>200$, reflecting the finite DOS at lower energies
around the hot spots.
In a three-dimensional (3D) model,
the relation $1/T_1 \propto T^{3}$ 
corresponds to the line-node SC gap.
If a point-node SC gap 
occurs in a 3D system owing to the present mechanism,
$1/T_1 \propto T^{5}$ will be observed.

Before closing this section, 
we comment that the mass enhancement factor 
$Z_\k$ is the function of the energy in the 
original Eliashberg equation.
In the present work, however,
we dropped the energy dependence for 
the simplicity of the analysis.
It will be allowed for the purpose of the present study, 
although it is not for calculating the fine
structure of the superconducting DOS.  
In Fig. \ref{fig:D0theta-Zw},
we show the solution of the Eliashberg equation
given by eqs. (\ref{eqn:Eliashberg}) and (\ref{eqn:Z})
as well as the solution given by taking the 
energy-dependence of $Z_\k(\w)$ into account correctly.
We see that the ratio of the gap anisotropy,
$\Delta_{\rm max}^0/\Delta_{\rm min}^0$, is enlarged a bit,
by performing a more accurate calculation.
In more detail,
$\Delta_{\rm max}^0$ is enlarged about 20\%
whereas $\Delta_{\rm min}^0$ 
(at the hot spot) increases only slightly,
because the influence of the SC gap function
on $Z_{\k}$ is little at the hot spot.
We have also checked that
the influence of the energy-dependence of $Z_\k$ 
on $T_{\rm c}$ is small.

\section{Comparison with the Weak Coupling BCS Theory}

In the previous section, 
we have shown that the s-wave superconductivity
with deep gap minima is reproduced easily 
by solving the strong-coupling Eliashberg equation
for $\lambda\equiv gN(0)=1.5$.
That is, a large anisotropic ratio 
$\Delta_{\rm max}/\Delta_{\rm min}>10$
is realized under the condition of s-$T_{\rm c}>$d-$T_{\rm c}$,
by assuming reasonable model parameters. 
In this section, we study the same model within the
weak-coupling BCS theory for a smaller value of $\lambda$, 
and show that the strong-coupling analysis performed 
in previous sections is indispensable for reproducing 
the strongly anisotropic s-wave SC state.

In the weak coupling BCS theory,
energy dependence of the gap function as well as
the normal self-energy are dropped.
Thus, the corresponding Eliashberg equation is 
given in eq.(\ref{eqn:Eliashberg}) 
(or eq.(\ref{eqn:Eliashberg-Tc}))
by (i) putting $\w=0$ ($\w_n=0$),
(ii) replacing $\Delta_\k(\e)$ ($\Delta_\k(\e_n)$) 
with $\Delta_\k^0$, and
(iii) putting $Z_\k=1$.
The obtained $\Delta_{\rm min,max}$ and s,d-$T_{\rm c}$ is
shown in Fig.\ref{fig:BCS},
which are denoted as ``BCS''.
Here, we put $\lambda=gN(0)=0.7$ to justify the weak coupling treatment.
We see that the s-$T_{\rm c}>d_{xy}$-$T_{\rm c}$ 
is realized only for $aN(0)<50$, where the anisotropy
of the s-wave SC gap is very weak.
Thus, the strong anisotropic s-wave SC gap cannot be obtained 
by the BCS theory.

Such a high d$_{xy}$-$T_{\rm c}$ given by the ``BCS'' theory
is suppressed drastically in the strong-coupling analysis,
mainly due to the mass-enhancement factor; $Z_\k>1$.
To show this fact, we also study the weak coupling BCS theory,
by taking the ``renormalization'' by $Z_\k$ into account correctly.
We call it the ``R-BCS'' theory,
where the Eliashberg equation is 
given in eq.(\ref{eqn:Eliashberg}) (or eq.(\ref{eqn:Eliashberg-Tc}))
by performing the simplification (i) and (ii) only.
The obtained $\k$-dependence of $Z_\k$ 
is shown in Fig.\ref{fig:Z}.
We see that $Z_\k$ takes large values around the hot spots,
so it will mainly reduce the effect of the AF fluctuations
in the Eliashberg equation.
The obtained result for $gN(0)=1.0$
by the R-BCS theory is shown in Fig. \ref{fig:BCS}.
We see that
d$_{xy}$-$T_{\rm c}$ is strongly reduced 
by the factor $Z_\k$, so
s-$T_{\rm c}>$d$_{xy}$-$T_{\rm c}$ 
is realized till much larger value of $aN(0)$; $aN(0)<240$.
Thus, $\Delta_{\rm max}/\Delta_{\rm min}\simle 5$
can be realized within the R-BCS theory.

As a result, 
the renormalization effect by $Z_\k$
is important to realize the highly anisotropic s-wave SC gap.
However, R-BCS theory is still insufficient
for a quantitative study,
so one have to work on the strong coupling Eliashberg
equation, eq. (\ref{eqn:Eliashberg}) or 
eq. (\ref{eqn:Eliashberg-Tc}), for $gN(0)\sim O(1)$.

Finally, we comment that the DOS around the hot spots
will decrease appreciably owing to the finite $\gamma_\k$
because $\gamma_\k$ takes a large value at the hot spots
when the AF fluctuations are strong,
as $Z_\k$ does.
This effect was found to reduce the $T_{\rm c}$
prominently in the case of high-$T_{\rm c}$ superconductors.
To take this effect into account,
we have to perform the $\e_\k$-integration ``numerically'' 
instead of using eq.(\ref{eqn:Gint}).
In this respect, the obtained d$_{xy}$-$T_{\rm c}$
in the present work may be over-estimated.

\section{Discussions}

\subsection{condition for the strongly anisotropic s-wave SC gap in 3D systems}

In the present paper,
we proposed a mechanism of the 
s-wave superconductivity with deep gap minima
in the presence of AF fluctuations.
We found that a 
highly anisotropic SC gap observed
in (Y,Lu)Ni$_2$B$_2$C can be reproduced 
with reasonable model parameters.
Here, we summarize the obtained condition for 
realizing the deep SC gap minima:

\noindent
(i) \underline{To make $\Delta_{\rm max}^0/\Delta_{\rm min}^0 \gg 1$};
a strong e-p coupling (e.g., $\lambda\equiv gN(0)>1$)
as well as larger $\xi_{\rm AF}^2$ (e.g., $\xi_{\rm AF}^2>50$)
is required. 
The ``radius of the minima'' in the SC gap
will be $\sim \xi_{\rm AF}^{-1}$. 
The strong coupling Eliashberg equation 
should be solved for a reliable analysis.

\noindent
(ii) \underline{To make s-$T_{\rm c}>$d-$T_{\rm c}$};
in addition to a strong e-p coupling,
a smaller coupling constant between electrons and AF fluctuations
(e.g., $aN(0)/\xi_{\rm AF}^2 <1$)
as well as a small energy scale of AF fluctuations
(e.g., $\w_{\rm sf}\ll\w_{\rm ph}$)
is required.
In addition, the nesting area of the FS should be small.
This condition will be easily satisfied in 
a three-dimensional system with several FS's.
This fact will be discussed later in more detail.

Here, the shape of the FS is chosen to be circle.
The most essential effect of the anisotropy as well as 
the dimensionality of the FS will appear in d-$T_{\rm c}$,
because it is enhanced when
the nesting of the FS which is
consistent with the main AF fluctuations exists,
while s-$T_{\rm c}$ and s-wave gap function will be rather insensitive.
Thus, the d-$T_{\rm c}$ obtained here
may have only a qualitative meaning.
Nonetheless, the realized d-$T_{\rm c}$ will 
remain small if one analyze a three dimensional (3D) system.
In fact, refs.
 \cite{Fukazawa,Arita,Nakamura,Monthoux}
discuss that the d-$T_{\rm c}$ due to AF fluctuations
in 3D systems remains low
because the nesting area in 3D FS is small in general.
In addition, the region of parameters for d-$T_{\rm c}$
is also very restricted in 3D systems.
According to a band calculation for YNi$_2$B$_2$C,
the shape of the FS is 3D like, and 
the nesting area of the FS is only 4.3\%,
which is supported experimentally 
by a tiny change of the resistivity at $T=T_{\rm N}$
 \cite{band-nesting}.

As a result,
d-$T_{\rm c}$ will not be high
in the real three dimensional FS for (Y,Lu)Ni$_2$B$_2$C,
so the mechanism of s-wave SC gap with deep minima
proposed in the present paper will be realized 
in (Y,Lu)Ni$_2$B$_2$C without difficulty.
In general, the weight of the nesting area in the FS decreases
further if several FS's exists like in heavy fermion systems.
Although the present study is based on a 
simplified two dimensional model,
the mechanism of the anisotropic s-wave SC gap 
due to AF fluctuations 
proposed in the present work will be universal.
We also not that the effect the finite 
$\gamma_\k={\rm Im}\Sigma_\k(-i\delta)$ at $T=T_{\rm c}$,
which will reduce d-$T_{\rm c}$ as mentioned 
in the previous section,
is not taken into account in the present analysis.
In this sense, the obtained d-$T_{\rm c}$ is overestimated.

Finally, 
we shortly discuss the 
location of the gap minima in 3D systems.
Figure \ref{fig:FS-3D} shows the
expected location of the gap minima
for (a)$|{\bf Q}|=2k_{\rm F}$
and (b)$|{\bf Q}|<2k_{\rm F}$
(${\bf Q}\parallel{\hat x}$ in both cases)
in the case of the spherical FS 
whose radius is $k_{\rm F}$.
In case (a),
point minima occur at two points on the $x$-axis.
In case (b), on the other hand,
gap minima will make two small circles.
Therefore,
gap minima will be short lines or small circles
in real 3D systems with anisotropic FS's.
Obviously, the depth of the gap minima
should depend on the position:
The deepest point will be
on the portion of the FS where $\lambda$ or
the density of states at $\k$,
Im$G_\k(-i\delta)/\pi$,
takes relatively a smaller value.
In such a case,
the deepest point of the gap may cause 
``point-node like behaviors''
in various thermodynamical quantities 
at lower temperatures,
even in case (b) ($|{\bf Q}|<2k_{\rm F}$).

For a more detailed study,
a strong coupling study based on a microscopic Hamiltonian
with 3D FS's is highly demanded.
For example, 
an analysis of a Holstein-Hubbard model 
by the FLEX approximation will be fruitful.

\subsection{possibility of the (s+id)-wave superconductivity:
breakdown of the time reversal symmetry}

In figs. \ref{fig:Delta0} and \ref{fig:Delta0-2},
we obtained s-$T_{\rm c}$ and d-$T_{\rm c}$
by considering the transition from the normal state.
In the case of s-$T_{\rm c} \gg $d-$T_{\rm c}$ 
(d-$T_{\rm c} \gg $s-$T_{\rm c}$),
a pure s(d)-wave SC state will be realized
whole the temperature region below s(d)-$T_{\rm c}$
because the established s(d)-wave SC gap will 
prevent the emergence of the d(s)-wave gap.
However, it is an interesting question
whether s-wave SC gap and d-wave one coexist or not
when d-$T_{\rm c} \sim $s-$T_{\rm c}$
below min$\{$d-$T_{\rm c}$, s-$T_{\rm c}\}$.
In this case,
the gap function will be given by
$\Delta_\k^0 = {\Delta_\k^^s}^0+e^{i\phi}{\Delta_\k^d}^0$,
where we choose the $U(1)$ gauge to make
both ${\Delta_\k^s}^0 \ (\propto 1)$ and 
${\Delta_\k^d}^0 \ (\propto k_x k_y)$ real.
Then, $e^{i\phi}=i$ is required
to make the gain of the condensation  
energy largest, at least within the weak-coupling theory,
because $\langle \Delta_\k^0 \rangle_{\rm FS}$
takes the maximum value then.
The ($s+id$)-wave superconductivity
is a very interesting state in that 
the time reversal symmetry (TRS) is broken.
In a later publication,
we will study the nature of the superconducting state
whole the temperature region in more detail
 \cite{future}.

Here, we notice that some of the deep minima on a
s-wave SC gap caused by the mechanism proposed in the 
present work might be filled to some extent
once the system changes from the s-wave SC state
to the ($s+id$)-wave state as the temperature decreases.
In the present model, in fact,
the position of the ``point minima'' in the anisotropic
s-wave state is different from that of the 
$d_{x^2-y^2}$-wave state except for the case of 
$|{\bf Q}|=2k_{\rm F}$,
as shown in Fig. \ref{fig:FS}.
We expect that a pure s-wave state is realized in 
(Y,Lu)Ni$_2$B$_2$C because the nesting area
of the FS is very small, which suggests that
$s-T_{\rm c}\ll d-T_{\rm c}$ as discussed before.

\subsection{consideration on other materials:
Sr$_2$RuO$_4$, PrOs$_4$Sb$_{12}$, Na$_{0.33}$CoO$_2$}

First, we stress that
the proposed mechanism here can make deep SC gap minima
also in a unconventional (p-wave or d-wave) superconductors
besides the original nodes:
When a unconventional superconductivity 
is realized by a Kohn-Luttinger 
type mechanism, which is free from the concept of the QCP,
some magnetic fluctuations owing to the nesting of the FS
will make ``deep gap minima'' on the SC gap function,
if their energy scale is too low to produce a different type
unconventional superconductivity.
Many heavy fermions and transition metal oxides superconductors 
have been reproduced properly in terms of the Kohn-Luttinger type 
mechanism, using the perturbation theory with respect to $U$
 \cite{Yamada-rev}.
In Appendix, 
we study a model which shows
the p-wave superconductivity with full gap,
and find that the deep gap minima
emerges on the SC gap function
as the AF fluctuations increases.

In a similar context, we would like to explain the 
theoretical study for Sr$_2$RuO$_4$ by Nomura and Yamada
 \cite{Nomura}:
Sr$_2$RuO$_4$ shows a spin-triplet SC state at $T_{\rm c}=1.5$K.
Based on the third-order and forth-order perturbation theory,
they found that the p-wave superconductivity 
occurs in Sr$_2$RuO$_4$.
According to their analysis,
a large and isotropic (chiral p-wave)
SC gap opens on the $\gamma$-FS, 
whereas small and strongly anisotropic gaps occur
on $\alpha$,$\beta$-FS's.
The obtained SC gap function
is consistent with the specific heat measurement 
below $T_{\rm c}$.
It is also consistent with the recent heat conductivity measurement
under the rotatable magnetic field
 \cite{Deguchi}.
We note that the almost gap-less SC gap realized in 
Sr$_2$RuO$_4$ originates from the cancellation of 
the pairing interaction.

PrOs$_4$Sb$_{12}$ is a Pr-filled Skutterudite superconducting
compound, with $T_{\rm c}=1.85$K
 \cite{Maple,Aoki}.
In PrOs$_4$Sb$_{12}$,
considerable magnetic (dynamical) fluctuations
are observed by $\mu$-SR measurement
 \cite{Aoki-muSR}
and by neutron diffraction measurement below 4K
 \cite{Kohgi}.
Moreover, e-p coupling will be also strong
in PrOs$_4$Sb$_{12}$ because 
various Pr-filled Skutterudite compounds
are conventional s-wave superconductors with relatively
high $T_{\rm c}$'s.

The symmetry of the SC gap is not determined now.
According to the heat conductivity measurement
under the rotatable magnetic field,
the SC gap has four point nodes along [100] and [010] axes
in a higher magnetic-field phase,
whereas two point nodes along [010] axis disappear
in a lower magnetic-field phase
 \cite{Izawa-Pr}.
One may expect that 
the strongly anisotropic s-wave SC state
is realized in PrOs$_4$Sb$_{12}$,
considering that 
the small energy scale of the AF fluctuations
observed by neutron diffraction ($\w_{\rm sf}\sim0.5$meV) 
 \cite{Kohgi}
is appropriate for making the deep SC gap minima
as shown in the present paper. 
Moreover, a recent $\mu$-SR measurement suggests that
the TRS is broken in the SC state of PrOs$_4$Sb$_{12}$
 \cite{Aoki-muSR}.
It may be a (s+id)-wave SC state
 \cite{Goryo},
which could be reproduced by the mechanism
proposed in the present work
as discussed in the previous subsection.

Na$_{0.33}$CoO$_2$ is a triangular lattice cobalt oxide 
superconductor, with $T_{\rm c}=4.5$K.
The symmetry of the SC gap is also under debate now.
According to NMR/NQR measurements,
$1/T_1T$ in a sample with magnetic fluctuations
shows no coherence peak, and $1/T_1\propto T^{3}$ is 
observed below $T_{\rm c}$
 \cite{Kyoto,Kitaoka}.
On the other hand, 
$1/T_1T$ in a sample with less magnetic fluctuations
shows a tiny coherence peak
 \cite{Sato}.
The reduction ratio of $T_{\rm c}$ due to impurities,
$-dT_{\rm c}/dx$ ($x$ being the concentration of impurities),
seems to be too small as a unconventional superconductor.
These measurements might be able to be explained 
as a s-wave superconductor with deep gap minima
caused by the mechanism proposed in the present paper,
as is the case with (Y,Lu)Ni$_2$B$_2$C.

In both PrOs$_4$Sb$_{12}$ and Na$_{0.33}$CoO$_2$,
more experimental and theoretical studies are required
to determine the symmetry and the mechanism of the superconductivity.
In future, the present study will offer a hint to understand 
various SC compounds in the presence of both 
e-p couplings and magnetic fluctuations.

\subsection{summary}

In summary,
we studied the influence of the AF fluctuations
on the SC gap function $\Delta_\k(\e)$ with s-wave symmetry
owing to the strong e-p coupling,
by analyzing the strong coupling Eliashberg equation.
We confirmed that deep SC gap minima emerge in $\Delta_\k(\e)$
as the AF fluctuation increases.
The condition for the realization of strong anisotropy
(say $\Delta_{\rm max}^0/\Delta_{\rm min}^0>10$)
is studied in detail.
We stress again that the main aim of this work
is to present the new mechanism of the point-node like 
SC gap, not to reproduce the precise value of $T_{\rm c}$
and thermodynamic measurements in (Y,Lu)Ni$_2$B$_2$C.

According to the present mechanism,
(groups of) pair of gap minima appear
at points on the Fermi surface which are
connected by the nesting vector ${\bf Q}$,
in both cases of s-wave superconductors
and non s-wave ones.
Their position is equal to the hot spots,
where the quasiparticle damping rate $\gamma_\k$
caused by the magnetic fluctuations takes 
the (local) maximum value.
The present mechanism is expected to reproduce the 
point-node like SC gap in (Y,Lu)Ni$_2$B$_2$C,
as well as the direction of the point minima.
It is a future problem to 
determine the precise positions of the gap minima
on the FS's by taking account of the realistic shape of FS's.

To realize a higher s-$T_{\rm c}$ ($\gg$d-$T_{\rm c}$),
it is desired that the position of the point minima
(hot spots) is away from the van-Hove singular point,
because $\lambda$ is expected to take a larger value there.
In case there are several hot spots,
the deepest point gap minima will appear 
on the FS where $\lambda$ takes the smallest value.
Deep SC gap minima in the SC gap due to AF fluctuations 
proposed in the present work are expected to emerge
even in unconventional superconductors,
as explained in Appendix.
In addition, recently found exotic superconductors,
PrOs$_4$Sb$_{12}$ and Na$_{0.33}$CoO$_2$,
were briefly discussed from the viewpoint of the
present study.


The author is grateful to Y. Matsuda, K. Izawa and J.P. Brison
for useful discussions about experiments
on (Y,Lu)Ni$_2$B$_2$C.
He is also grateful to M. Sato, H. Yoshimura and K. Ishida
for valuable comments on NMR/NQR.
For useful discussions on theory,
he is thankful to
K. Yamada, T. Sato, H. Kohno, H. Ikeda and T. Nomura.

\appendix
\section{In the case of the p-wave superconductor}

As discussed in \S VI,
the deep SC gap minima
at the hot spots owing to the AF fluctuations
proposed in the present paper
will also be realized 
in the case of p- or d-wave superconductor.
To confirm this fact,
we study a p-wave superconducting systems
with AF fluctuations, and see that 
the deep SC gap minima emerge
as the AF fluctuations increases.
In general,
the odd-parity SC gap function in the Nambu representation
is given by
\begin{eqnarray}
{\hat \Delta}_\k= {\bf d}_{\k}\cdot i{\vec \s}\s_y,
\end{eqnarray}
where ${\bf d}_{\k}$ is the d-vector.
The quasiparticle energy gap is given by 
$\sqrt{{\bf d}_\k^\ast \cdot {\bf d}_\k}$.

Here,
we introduce the following paring potential 
for the p-wave channel in two dimension:
\begin{eqnarray}
V_{\k,\k'}^{\mbox{p-ch}}(\w_n)
 &=& -g\frac{\w_{\rm p}^2}{\w_n^2+\w_{\rm p}^2} 
 \cdot 2\cos(\theta_\k-\theta_{\k'}) ,
 \label{eqn:ap2}
\end{eqnarray}
whose diagrammatic expression 
is shown in Fig. \ref{fig:p-channel}.
In the presence of the AF fluctuations,
the system has the tetragonal symmetry ($D_{4h}$).
In this case, 
six $p$-wave pairing states with different symmetries
give the same quasiparticle gap without nodes,
which is expected to optimize the condensation energy
at least within the weak coupling theory
 \cite{Sigrist}.
In each pairing state,
${\bf d}_\k(\w)$ is composed of 
($\Delta_\k^{(1)}(\w), \Delta_\k^{(2)}(\w)$),
which changes under the point group operations
as (${\hat k}_x, {\hat k}_y$).
We promise that both
$\Delta_\k^{(1)}(0)$ and $\Delta_\k^{(2)}(0)$ are real.
For example, 
the chiral p-wave state is given by
${\bf d}_\k(\w)\equiv {\hat {\bf z}}
 (\Delta_\k^{(1)}(\w) \pm i \Delta_\k^{(2)}(\w))$.
Then, the strong coupling Eliashberg equation 
for the p-wave symmetry with full gap is given by
\begin{eqnarray}
\Delta_\k^{(j)}(\w_n) &=&
 -T\sum_{m}\frac{\pi N(0)}{Z_\k}
 \int_{\rm FS}\frac{d\Omega_\p}{2\pi}
 \frac{\Delta_\p^{(j)}(\w_m)}{\sqrt{\w_m^2+D_\p^2(\w_m)}}
  \nonumber \\
& &\times \left(V_{\k,\p}^{\mbox{p-ch}}(\w_n-\w_m)
 +\frac{-1}{3}V_{\k-\p}^{\rm AF}(\w_n-\w_m)
 \right) ,
 \label{eqn:ap1} 
\end{eqnarray}
where $j=1,2$, 
$D_\p^2(\w) \equiv 
\{\Delta_\p^{(1)}(\w)\}^2 + \{\Delta_\p^{(2)}(\w)\}^2$, 
and $V_{\k-\p}^{\rm AF}(\w_n)$ 
represents the interaction due to AF fluctuations
introduced in eq.(\ref{eqn:VAF}).
Note that the factor $-1/3$ in front of $V_{\k-\p}^{\rm AF}$
comes from the fact that
$\langle{\vec s}\cdot{\vec s'}\rangle_{\mbox{S=1}}
=(-1/3)\langle{\vec s}\cdot{\vec s'}\rangle_{\mbox{S=0}}=1/4$.
Because of the relation
$\cos(\theta_\k-\theta_{-\k})=-1$, 
the expression for $Z_\k$ in eq.(\ref{eqn:ap1}) is 
same as that in eq.(\ref{eqn:Z}).

\begin{figure}
\begin{center}
\epsfig{file=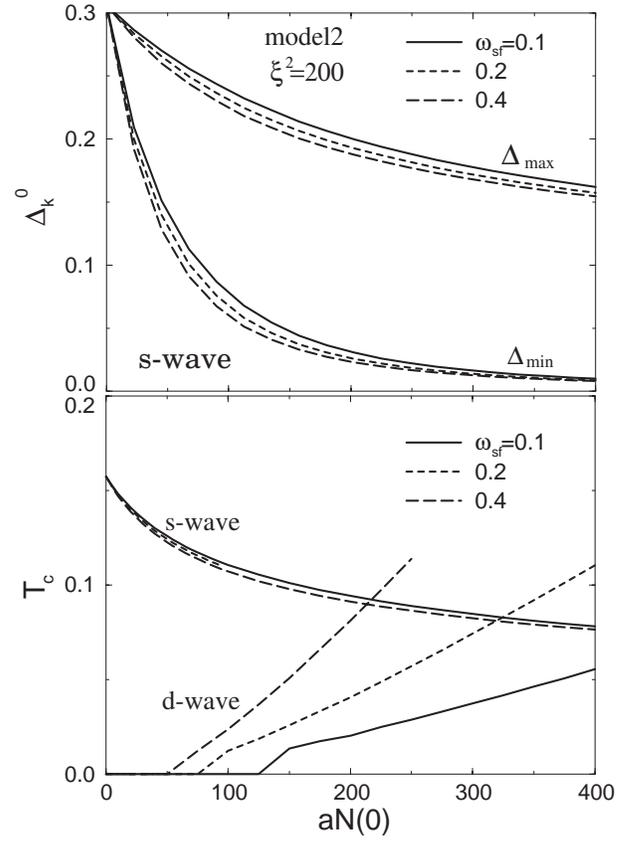,width=8cm}
\end{center}
\caption{
$\Delta_{\rm min, max}^0$ and s,d-$T_{\rm c}$
obtained in model 2 for $\xi_{\rm AF}^2=200$.
}
  \label{fig:Delta0-2}
\end{figure}

\begin{figure}
\begin{center}
\epsfig{file=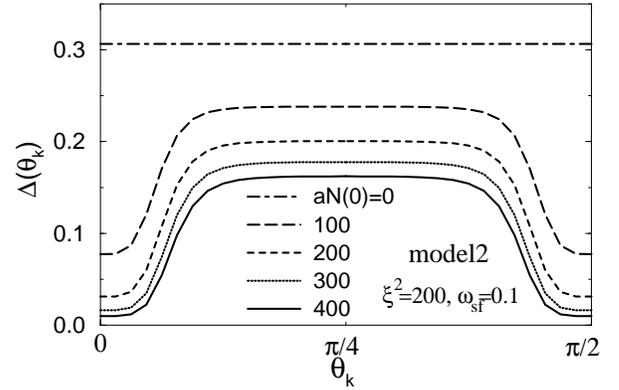,width=8cm}
\end{center}
\caption{
$\theta_\k$-dependence of $\Delta_{\k}^0$
obtained in model 2.
}
  \label{fig:D0theta}
\end{figure}

\begin{figure}
\begin{center}
\epsfig{file=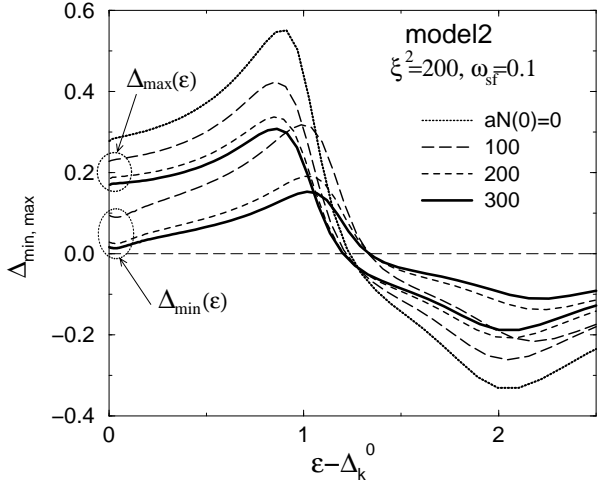,width=8cm}
\end{center}
\caption{
$\e$-dependence of $\Delta_{\rm min, max}(\e)$
obtained in model 2.
Here we put $\Gamma=\w_{\rm sf}/6$.
}
  \label{fig:Gap-w}
\end{figure}

\begin{figure}
\begin{center}
\epsfig{file=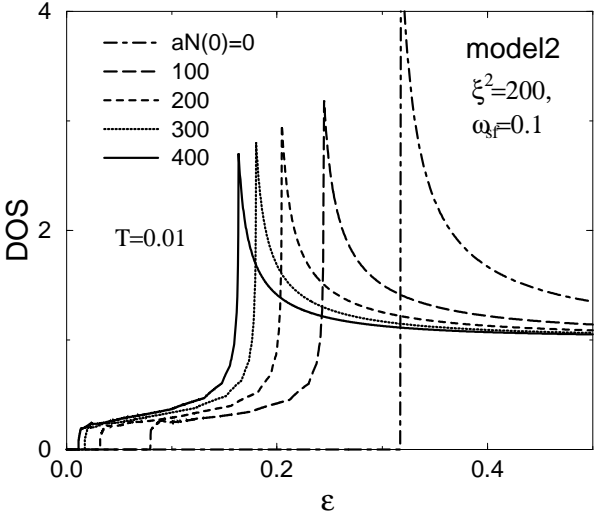,width=8cm}
\end{center}
\caption{
$\e$-dependence of $\rho(\e)$
obtained in model 2. We put $N(0)=1$.
}
  \label{fig:DOS}
\end{figure}

\begin{figure}
\begin{center}
\epsfig{file=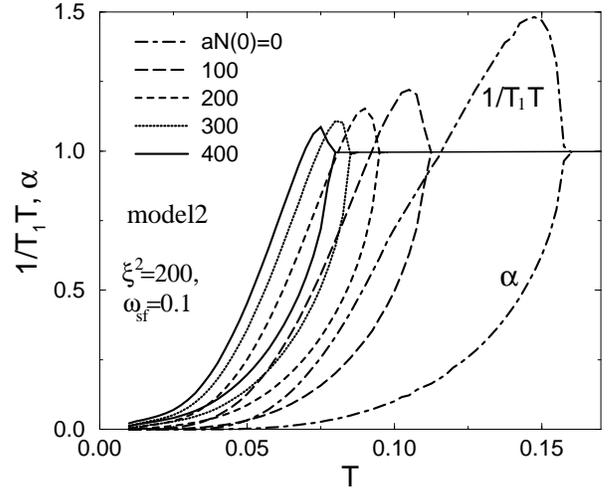,width=8cm}
\epsfig{file=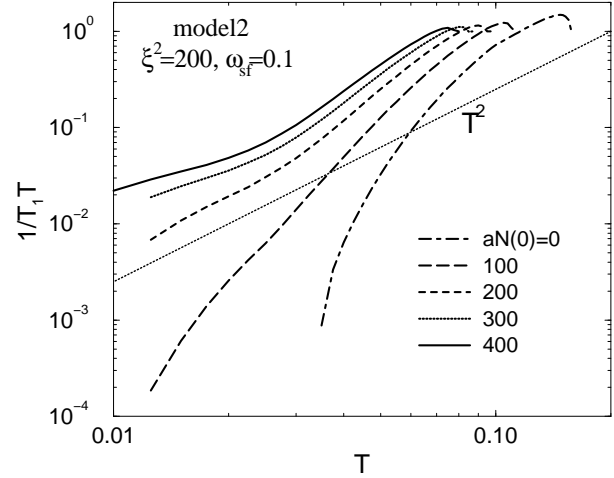,width=8cm}
\end{center}
\caption{
(a) Obtained nuclear-lattice relaxation ratio $1/T_1T$ 
and ultrasonic attenuation ratio $\alpha$.
(b) $1/T_1$ is proportional to $T^3$ at lower temperatures
for $aN(0)>200$, reflecting the deep minima in SC gap.
$1/T_1 \propto T^5$ should be observed in a three-dimensional model. 
}
  \label{fig:T1T}
\end{figure}

\begin{figure}
\begin{center}
\epsfig{file=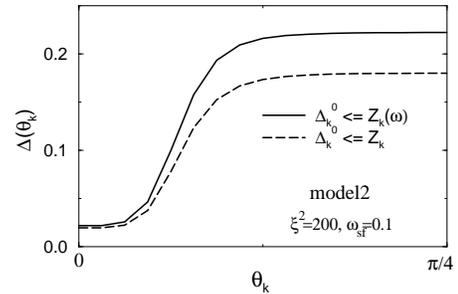,width=6cm}
\end{center}
\caption{
$\theta_\k$-dependence of $\Delta_{\k}^0$
obtained by solving Eliashberg equation when $aN(0)=300$,
(i) by dropping the energy-dependence of $Z_\k$
(dashed line) and (ii) by taking the 
the energy-dependence of $Z_\k(\w)$ into account
correctly (full line).
}
  \label{fig:D0theta-Zw}
\end{figure}

\begin{figure}
\begin{center}
\epsfig{file=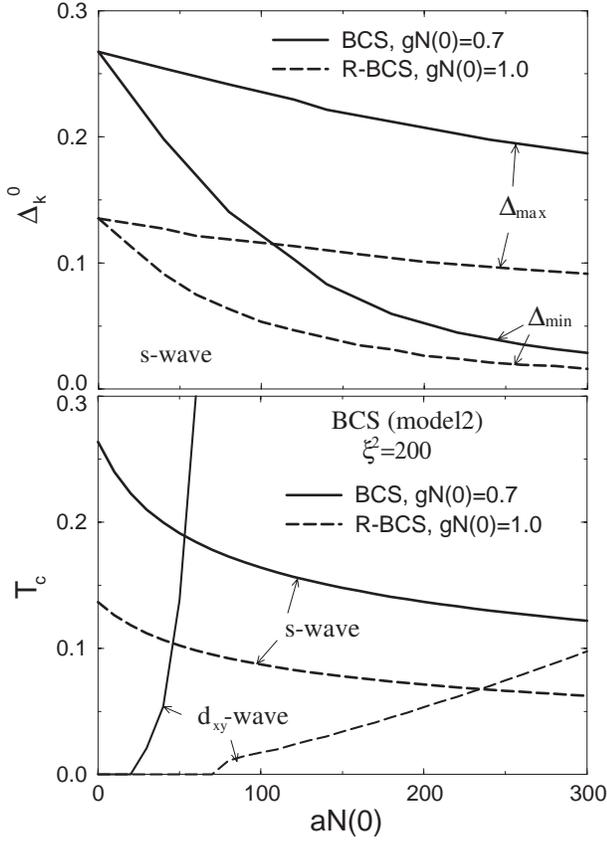,width=8cm}
\end{center}
\caption{
Obtained $\Delta_\k^0$ and s,d-$T_{\rm c}$
by BCS theory ($Z_\k=1$) and by R-BCS theory
($Z_\k>1$), respectively.
}
  \label{fig:BCS}
\end{figure}

\begin{figure}
\begin{center}
\epsfig{file=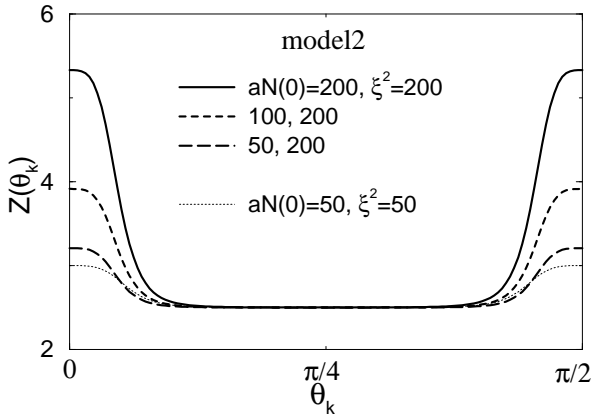,width=8cm}
\end{center}
\caption{
$\theta_\k$-dependence of $Z_\k$ for $\lambda=1.5$. 
}
  \label{fig:Z}
\end{figure}

\begin{figure}
\begin{center}
\epsfig{file=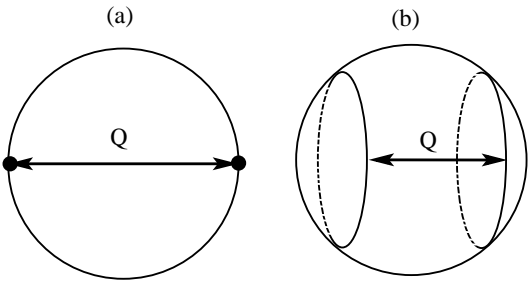,width=7cm}
\end{center}
\caption{
Illustration of the 
location of the gap minima for 
(a)$|{\bf Q}|=2k_{\rm F}$ and 
(b)$|{\bf Q}|<2k_{\rm F}$ (${\bf B}\parallel{\hat x}$),
in the case of the spherical FS.
}
  \label{fig:FS-3D}
\end{figure}

\begin{figure}
\begin{center}
\epsfig{file=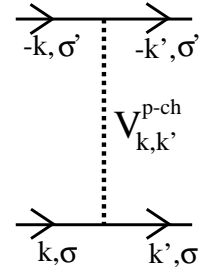,width=2.5cm}
\end{center}
\caption{
Paring interaction for the p-wave channel.
}
  \label{fig:p-channel}
\end{figure}

\begin{figure}
\begin{center}
\epsfig{file=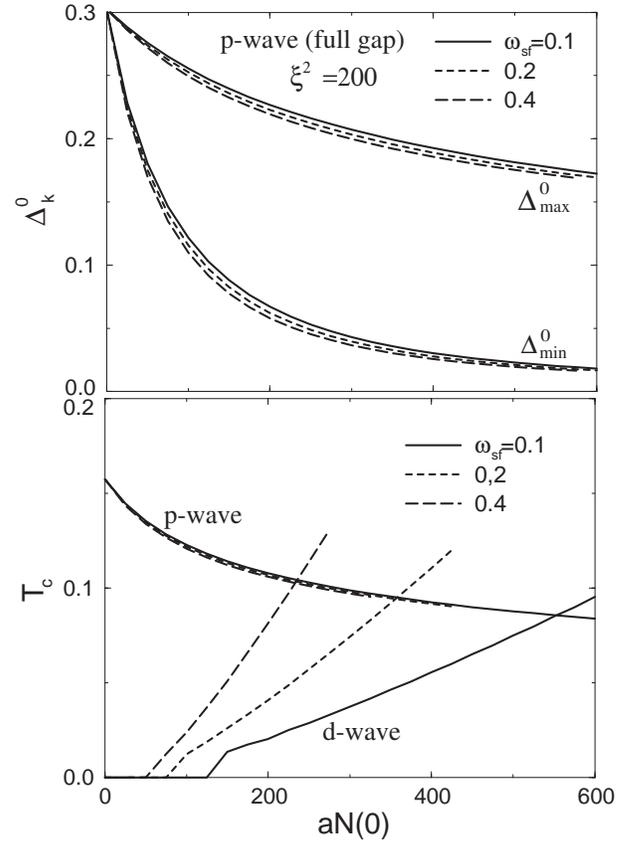,width=8cm}
\end{center}
\caption{
Obtained quasiparticle gap function
$\Delta_\k^0$ for the full-gap p-wave SC state
and $T_{\rm c}$'s for both p- and d-wave solutions,
as functions of $aN(0)$.
}
  \label{fig:p-DTc}
\end{figure}

Here, we study eq.(\ref{eqn:ap1}) numerically,
by putting $g= \w_{\rm p}= 1.0$.
The quasiparticle gap is given by the relation
$\Delta_\k^0 \equiv |D_\k(\Delta_\k^0)|$.
The upper panel in Fig. \ref{fig:p-DTc}
shows the obtained 
$\Delta_{\rm max}^0 \equiv \Delta_{\theta_\k=0}^0$ and
$\Delta_{\rm min}^0 \equiv \Delta_{\theta_\k=\pi/4}^0$
for the p-wave SC state at zero temperatures,
as functions of the AF coupling constant $aN(0)$.
(Note that $|\Delta_\k|$ is constant when $aN(0)=0$.)
The lower panel shows the obtained
$T_{\rm c}$'s both for p-wave and d-wave.
As shown in Fig. \ref{fig:p-DTc},
deep minima emerges in the p-wave SC gap
owing to the AF fluctuations.
Quantitatively speaking, however,
the obtained ratio $\Delta_{\rm max}/\Delta_{\rm mim}$
in the present model is smaller
than that in Fig. \ref{fig:Delta0-2}
at a fixed $aN(0)$,
because of the factor $1/3$ in front of
$V_{\rm AF}$ in eq.(\ref{eqn:ap1}).



\end{multicols}

\end{document}